\documentclass[aps,showpacs,superscriptaddress,amsfonts,prfluids]{revtex4} 

\usepackage{graphicx}
\usepackage{array}
\usepackage{hyperref}
\usepackage{amsmath,amssymb}
\usepackage{bm}
\usepackage{float}
\usepackage{amsfonts}
\usepackage{xcolor}
\usepackage{ar}

\begin{document}
\title{Geometric phase of spheroidal particles in a fluid flow and its control}
\date{\today} 

\author{Jorge Arrieta}
\affiliation{Instituto Mediterr\'aneo de Estudios Avanzados, IMEDEA, UIB-CSIC, Esporles, 07190, Spain}
\author{Julyan H. E. Cartwright}
\affiliation{Instituto Andaluz de Ciencias de la Tierra, CSIC--Universidad de Granada, 18100 Armilla,
Granada, Spain} 
\affiliation{Instituto Carlos I de F\'{\i}sica Te\'orica y Computacional, Universidad de Granada, 18071 Granada, Spain}
\author{Oreste Piro}
\affiliation{Instituto Mediterr\'aneo de Estudios Avanzados, IMEDEA, UIB-CSIC, Esporles, 07190, Spain}
\affiliation{Departament de F\'{\i}sica, Universitat de les Illes Balears, 07122, Palma de Mallorca, Spain} 
\author{Idan Tuval}
\affiliation{Instituto Mediterr\'aneo de Estudios Avanzados, IMEDEA, UIB-CSIC, Esporles, 07190, Spain}
\affiliation{Departament de F\'{\i}sica, Universitat de les Illes Balears, 07122, Palma de Mallorca, Spain}

\begin{abstract}
We investigate the dynamics of spheroids immersed in the journal bearing flow subject to a contractible non-reciprocal loop. We show how geometric phases appear not only in the position, but also in the orientation of such particles. We show how control and targeting can be carried out on spheroidal particles in the flow.
\end{abstract}

\maketitle

\section{Introduction}

The geometric phase is a beautiful example of the generality of physics. If one takes a system around a closed circuit in its parameters, then it is possible that some variables may not return to their initial values. This was first pointed out for a quantum system by Berry\cite{Berry1984}, but was straight away understood to be just as applicable in the classical world\cite{shapere}.  Following its discovery in the 1980s, earlier examples and classical cases in which a geometric phase appears have been pointed out from all across physics and beyond. In fluid mechanics, the geometric phase is involved in being able to swim at low Reynolds number \cite{shapere2}. We have shown that it can be used in mixing at low Reynolds number \cite{Arrietaetal2015,Arrietaetal2020}, and in particular, this geometric mixing can explain some aspects of how mixing occurs in the stomach during the digestive process, where it is peristaltic movements of the stomach walls that permit mixing \cite{Arrietaetal2015}.

In this paper we investigate the advection of spheroidal particles induced by taking closed circuits in the parameters defining a two-dimensional journal-bearing flow. Besides the transport in space caused by an accumulated geometric phase, the rotation of anisotropic particles results in an additional angular phase at the completion of each parameter's cycle. We analyse the dynamics of this angular phase as a function of the controlling parameters with a focus on the role of the eccentricity of the inner cylinder. We show that a finite angular phase emerges even in the limit of small eccentricity. We further use AI techniques to illustrate how the geometric phase can be exploited to drive particles to specific targets in Stokes flows. Machine learning has emerged recently as an effective tool to study the response of active particles to external cues in unbounded flows, optimize their navigation strategy to maximize distance covered \cite{Colabrese2017}, to approach regions of high vorticity \cite{Colabrese2018} or to evade other microswimmers by active cloaking \cite{Mirzakhanloo2020}. However, using reinforcement learning algorithms to optimize the transport of passive particles in bounded creeping flows remains largely unexplored.

\section{Formulation}

We consider the advection and the rotation of a prolate spheroid induced by the journal bearing flow in the viscous limit. The journal bearing flow can be described analytically in bipolar coordinates ($\xi,\eta$)~\cite{BallalRivlin}, whereas the motion of the non-inertial spheroid is obtained integrating
\begin{align}
\frac{d \xi}{d t}&=\frac{1}{H}u_\xi \label{uxi}\\
\frac{d \eta}{d t}&=\frac{1}{H}u_\eta \label{ueta}
\end{align}   
where $u_\xi$ and $u_\eta$ represent the components of the velocity in bipolar coordinates with $H=b/(\cosh \xi -\cos \eta)$ the scale factor and $b$ the focus of the coordinate system.

The hydrodynamics of ellipsoidal particles at low Reynolds numbers was comprehensively investigated by Jeffery~\cite{Jeffery1922}. He showed that, in order for the mean torque acting on the particle to vanish, the particle in general must rotate. For spheroids in two-dimensional flows, as considered herein, the dynamics is further simplified since two semi-axis are equal and the rotation of the particle only occurs in the axis perpendicular to the plane of the flow. Hence, the rate of rotation is simply given by 
\begin{equation}
\omega=1/2\nabla\times\mathbf{u}+(\AR^2-1)/(\AR^2+1)\varepsilon_{x'y'},
\label{viscoustorque}
\end{equation}
where $\AR$ is the aspect ratio of the particle and $\varepsilon_{x'y'}$  is the component of the strain rate tensor referred to the local coordinates of the spheroid with $x'$ denoting the axis of revolution. Moreover, the instantaneous angle of the particle $\varphi$, as measured from the angle between $x'$ and the horizontal, can be obtained integrating $d\varphi/d t=\omega$.

\section{Results}

\begin{figure*}[t!]
\begin{center}
\includegraphics*[width=1.0\textwidth]{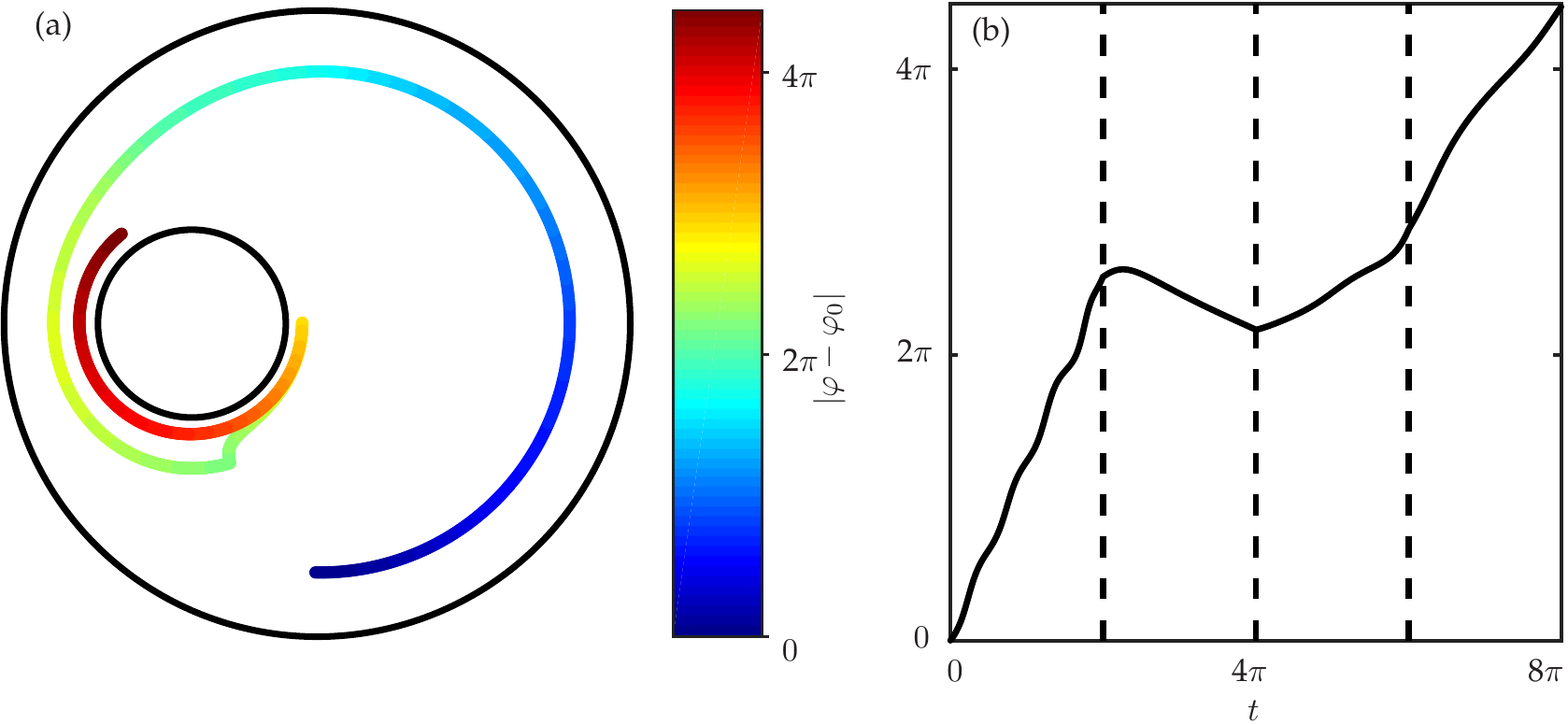}
\end{center}
\caption{ (a) The trajectory of a spheroidal particle of aspect ratio $\AR=10$ within a journal bearing flow. The radii of the 
outer cylinder and the inner cylinder are $R_1=1$ and $R_2=0.3$ respectively and the eccentricity parameter is $\varepsilon=0.4$. 
Along a closed loop each cylinder is rotated $2\pi$ counterclockwise and clockwise, yielding a loop of area $2\pi\times2\pi$. The 
colormap plotted along the trajectory represents the modulus of the angle rotated by the particle measured from the initial orientation 
$\varphi_0=0$. (b) The temporal evolution of the modulus of the angle rotated by the particle, $|\varphi-\varphi_0|$, measured 
from the initial orientation.}
\label{fig:fig1}
\end{figure*}

If one takes a closed loop enclosing a non-zero area in parameter space for the journal-bearing flow - for instance, by rotating in a non-contractible fashion the inner and outer cylinders back and forth by a finite angle - a geometric phase arises in the position of any passive scalar~\cite{Arrietaetal2015}. Instead, in the work reported here, we find that if we look at the dynamics of a spheroidal particle in the flow, the non-contractible cycle also induces an angular phase which results in a difference between the initial and the final orientation of the particle at the end of the same loop. We illustrate this effect in Figure~\ref{fig:fig1} where we show the integrated trajectory of a spheroidal particle after a single cycle of area $2\pi\times 2\pi$ (i.e., with both cylinders rotated through $2\pi$). As expected, a geometric phase arises taking the particle not to retrace its trajectory when the boundaries return to their initial position. In addition, the angle advanced by the particle with respect to its initial in-plane orientation is illustrated by the colour map along the trajectory - in panel (a) -, and by the temporal evolution of the angle - in panel (b) - , showing clearly that the particle fails to return to its original orientation.

\begin{figure*}[t!]
\begin{center}
\includegraphics*[width=1.0\textwidth]{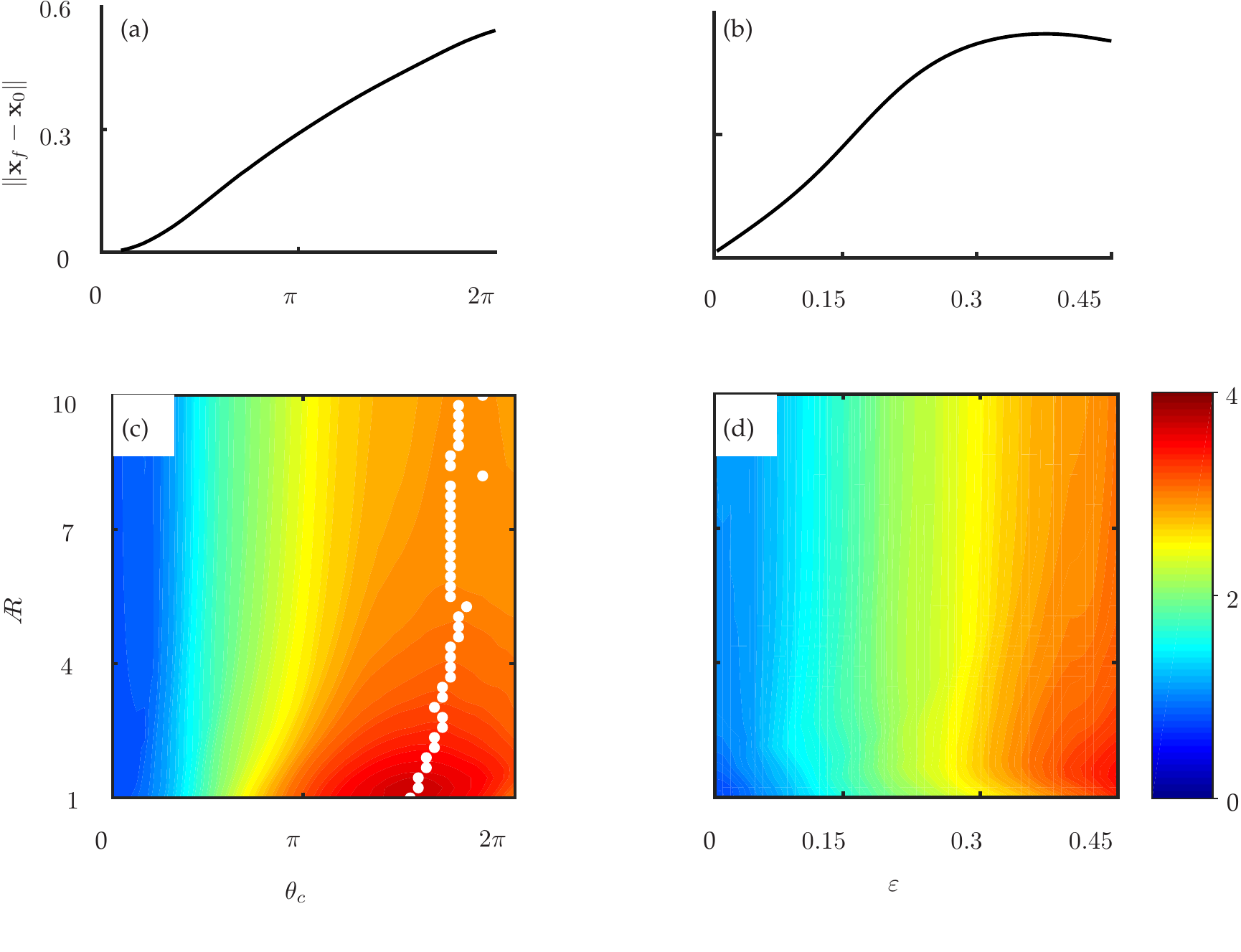}
\end{center}
\caption{ The effect of the aspect ratio, the area of the loop and the eccentricity of the inner cylinder in the phases of the 
spheroid. The geometric phase is measured as the Euclidean distance of the final position and the initial position of the particle 
after one cycle whereas the angular phase of the spheroid is defined as the modulus of the angle rotated by the particle from its initial orientation, $|\varphi-\varphi_0|$. In both sweeps the radius of the 
outer and the inner cylinder were set to $R_1=1$ and $R_2=0.3$ respectively. (a) The geometric phase  of the particle as a function of 
the angle rotated by each cylinder $\theta_c$ for $\varepsilon=0.4$. (b)The geometric phase as a function of eccentricity for
an area of the loop $2\pi\times 2\pi$. (c) The angular phase as a function of the aspect ratio of the spheroid and the angle rotated by each cylinder $\theta_c$ 
measured from the initial orientation after one loop as a function of the aspect ratio
and the area of the loop. White dots represent the maximum angular phase for a given aspect ratio and illustrate the non-monotonic dependence  of the angular phase with the area of the loop, $\theta_c\times\theta_c$.(d) The angular phase as a function of the aspect ratio and the area of the 
loop.In (a)-(d) the data depicted are obtained by averaging the geometric and the angular phase over $\sim 6.4\cdot 10^4$  trajectories 
of particles distributed uniformly within the domain.}
\label{fig:fig2}
\end{figure*}

We further explore the effect of the aspect ratio of the spheroid, $\AR$, the area of the loop, and the eccentricity of the position of the inner cylinder, $\varepsilon$, on both the geometric and the angular phases by performing two parametric sweeps at fixed radii of both cylinders: first, by fixing the eccentricity $\varepsilon=0.4$ and studying the coupled effect of the area of the cycle and $\AR$; second, by fixing the area of the loop to $2\pi\times2\pi$ and exploring instead the $(\varepsilon,\AR)$ plane. Figure~\ref{fig:fig2} shows the results for the geometric phase (a-b) and the angular phase (c-d) on the different parameters. As expected, in all cases the effect of the aspect ratio of the particle only affects the angular phase. For small areas of the loop, the angular phase remains small for any $\AR$ as it is mostly controlled by the vorticity of the flow in the limit of nearly spherical particles (i.e., where the term ($\AR^2-1)/(\AR^2+1)$ becomes almost negligible). For increasing values of the area of the loop, the angular phase shows a non-monotonic behaviour with the aspect ratio of the particle with distinct maxima for large areas. A similar non-monotonic behaviour is observed for the angular phase as a function of $\varepsilon$. It is worth noting that the local maxima for the geometric and angular phases do not overlap.

\begin{figure*}[t]
\begin{center}
\includegraphics*[width=1.0\textwidth]{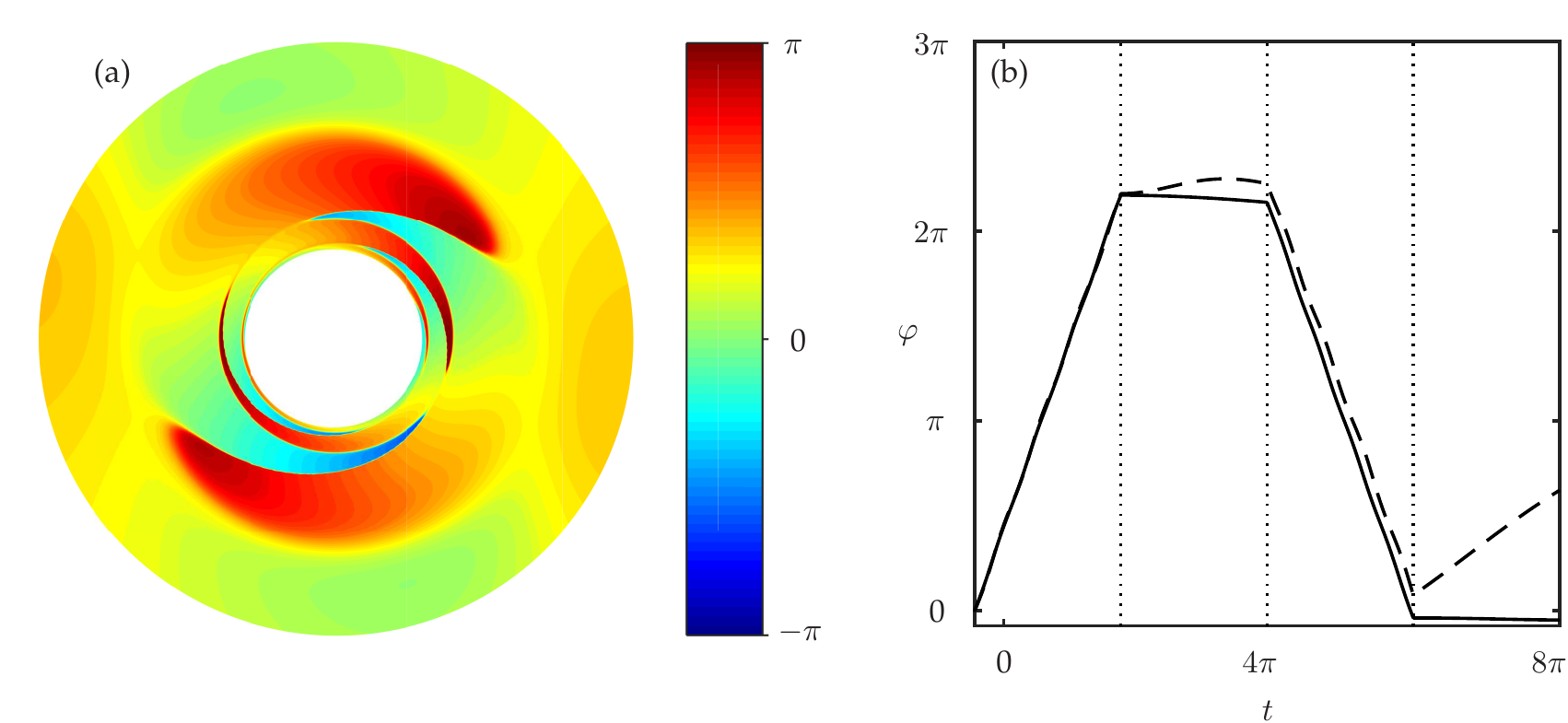}
\end{center}
\caption{(a) The spatial distribution of the angular phase for $\varepsilon=0.01$ and $\AR=2.35$. (b) Two different trajectories 
for a single loop departing from the inner part of the cylinder $(x_0,y_0)=(0,0.6)$ (dashed line) and the outer part of the cylinder 
$(x_0,y_0)=(0,0.8)$ (solid line) }
\label{fig:fig4}
\end{figure*}
The performed parameter sweeps highlight that combinations of the parameters that yield a small geometric phase also yield a non-negligible angular phase. This is clearly apparent in figure~\ref{fig:fig2}(d) where even small eccentricities and small aspect ratios yield a non-zero angular phase. This effect can be mostly attributed to the rotation induced by the strain rate, since in that limit the vorticity recovers its value at the beginning of the loop and, hence, the rotation it induces remains small. To illustrate the non-trivial dynamics of the angular phase, in figure~\ref{fig:fig4}(a) we have represented the spatial dependence of the angle rotated by the spheroid for a set of values in the lower left corner of figure~\ref{fig:fig2}(d). As can be seen, the angle rotated by the particle is non-zero in a large portion of the domain, ensuring that the average of the angular phase is of order one, even in the limit of the eccentricity tending to zero. The heterogeneity of the angular phase is further illustrated in figure~\ref{fig:fig4}(b), where the evolution of the angle of two particles departing from nearby initial conditions (e.g. close to the wall of the inner cylinder) can have an almost zero angular phase (solid line) or an angular phase close to $\pi$ (dashed line).

These results suggest a new kind of irreversibility for spheroidal particles, in which a particle returns to its original position while failing to get back to its initial orientation. We can rationalize this result by considering the limiting case of two concentric cylinders, $\varepsilon=0$. In this configuration the flow-field can be simply described in terms of polar coordinates $(r,\theta)$, where the polar velocity is $u_\theta=\pm\Omega_1/[1-(R_2/R_1)^2](r-R_2^2/r)$ while the outer cylinder is rotated with an angular velocity $\Omega_1$ and the inner cylinder is at rest (i.e., $ t \in[0, 2\pi]$ and $t\in[4\pi, 6\pi]$), whereas $u_\theta=\pm\Omega_2/[1-(R_2/R_1)^2](1/r-r/R_1^2)$ is the polar velocity when the inner cylinder is rotated and the outer remains at rest (corresponding to $t\in[2\pi, 4\pi]$ and $t\in[6\pi, 8\pi]$ for the selected loop). Note that in both expressions the sign of the velocity field represents the direction in which the cylinder is being moved along the loop (the positive sign holding for $t\in[0,2\pi]$ and $t\in[2\pi,4\pi]$ and the negative for $t\in[4\pi,6\pi]$ and $t\in[6\pi,8\pi]$). These simplified flow fields allow us to obtain a simpler version of the differential equation that determines the angle rotated by the spheroid, that takes the following form:
\begin{equation}
\frac{d\varphi}{d t}=\Lambda+\Gamma\cos[2(\theta+\varphi)],
\label{eq:eq_conc}
\end{equation}
with $\Lambda=\pm\Omega_1/[1-(R_2/R_1)^2][2-(R_2/r)^2]$ and $\Gamma=\pm\Omega_1/[1-(R_2/R_1)^2](\AR^2-1)/(\AR^2+1)(R_2/r)^2$ when the outer cylinder is rotated ($\Omega_1=\pm 1$ and $\Omega_2=0$), whereas $\Lambda=\pm\Omega_2/[1-(R_2/R_1)^2][(R_2/r)^2-2(R_2/R_1)^2]$ and $\Gamma=\mp\Omega_2/[1-(R_2/R_1)^2](\AR^2-1)/(\AR^2+1)(R_2/r)^2$ when $\Omega_1=0$ and $\Omega_2=\pm 1$. The same criterion holds for the signs in the expressions above. 
 
\begin{figure*}[t]
\begin{center}
\includegraphics*[width=1.0\textwidth]{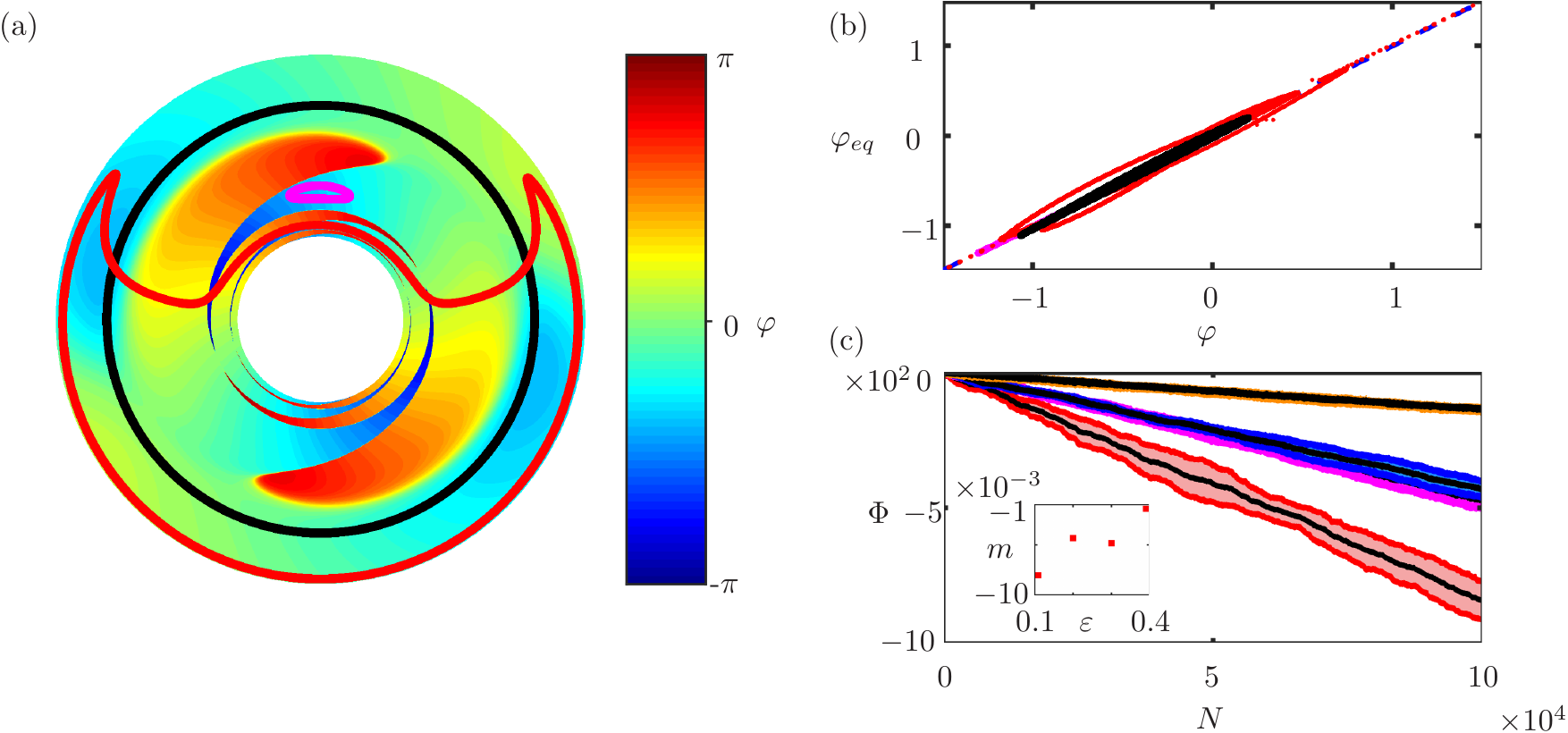}
\end{center}
\caption{(a) The spatial distribution of the equilibrium angle $\varphi_{eq}$ 
by a spheroid of $\AR=10$ after one cycle for the concentric cylinder case. 
The value of $\varphi_{eq}$ has been obtained by integrating~\eqref{eq:eq_conc} for a 
loop of $2\pi\times 2\pi$. Black line shows a periodic trajectory in the case $\varepsilon=10^{-5}$, whereas the trajectories depicted in magenta and red correspond to the case $\varepsilon=10^{-2}$. (b) The relation between the equilibrium angle $\varphi_{eq}$ evaluated at the location of the three trajectories plotted in (a) and the orientation of the spheroid obtained by direct integration. In the three cases deviations from a line of slope one (dashed blue line) and in consequence from $\varphi_{eq}$ are small. (c) The cumulative angle for $\varepsilon=0.1$(red), 0.2 (blue), 0.3 (magenta) and 0.4 (orange), retaining only significant variations in the orientation between two consecutive cycles. For each eccentricity trajectories departing from $x_0=0$ and $y_0=0.32,0.4,0.6,0.8,0.9$ were integrated for 5$\times 10^{5}$ loops. The average cumulative angle is represented for each eccentricity by the black solid line, whereas dark lines of the corresponding colour correspond to the sum and the difference of the average angle and the standard deviation. The slope $m$ of the linear fitting of the average cummulative angle is represented in the inset.}
\label{fig:fig5}
\end{figure*}

Both the numerical integration and the analytical solution of the simplified differential equation reveal that the orientation of the spheroid becomes independent at leading order of the initial orientation after the first rotation of the inner cylinder. Therefore, for a given radial position after one loop the orientation of the particle becomes independent of its initial value. In consequence, further repetitions of the rotation protocol end in the same orientation and no angular phase is gained after the first cycle. However, this equilibrium angle is dependent on the position within the flow, as can be seen in the contour map of Fig.~\ref{fig:fig5}(a), where the spatial distribution of the orientation of the spheroid after one closed loop is shown. For small eccentricities particles perform periodic trajectories and the orientation of the particle is a small perturbation of the equilibrium orientation $\varphi_{eq}$ for $\varepsilon=0$. Fig.~\ref{fig:fig5}(a) further depicts three trajectories with different initial conditions and for $\varepsilon=10^{-5}$ (black) and $10^{-2}$ (red and magenta). We compare their equilibrium angle at the location of the particles and the angle rotated by the spheroids obtained by direct numerical integration in Fig.~\ref{fig:fig5}(b). It can be seen that the angle rotated by the particle is just a periodic perturbation around the equilibrium angle of the concentric solution.

As the eccentricity of the journal-bearing flow increases, the observed periodic behaviour breaks and the angle rotated by the spheroids randomises while remaining bounded. A detailed analysis of the long-term dynamics ($~10^{5}$ cycles) of these trajectories clearly shows that during most of the cycles the variation of the rotated angle is very small with only a few cycles where it changes by an order of magnitude. Hence, the orientational dynamics of the particles resembles a L\'evy flight \cite{Klafter1996}. We took advantage of this separation of scales to show that  an effective gain in the angle rotated by the particle is indeed produced after each of these jumps. Based on the distributions of the difference of the angle between two consecutive loops, which show a peaked core with long tails, we operationally define a L\'evy jump as an event where the gain between two consecutive loops is larger than an angular threshold of $0.1$. We show in Fig.~\ref{fig:fig5}(c) the average (solid black lines) and standard deviation (solid coloured lines) of the effective gain in the rotated angle for different values of $\varepsilon$. Moreover, the slopes of these curves show a weakly non-monotonic behaviour with $\varepsilon$.

\section{Active control of spheroidal particles}

\begin{figure*}[t]
\begin{center}
\includegraphics*[width=1.0\textwidth]{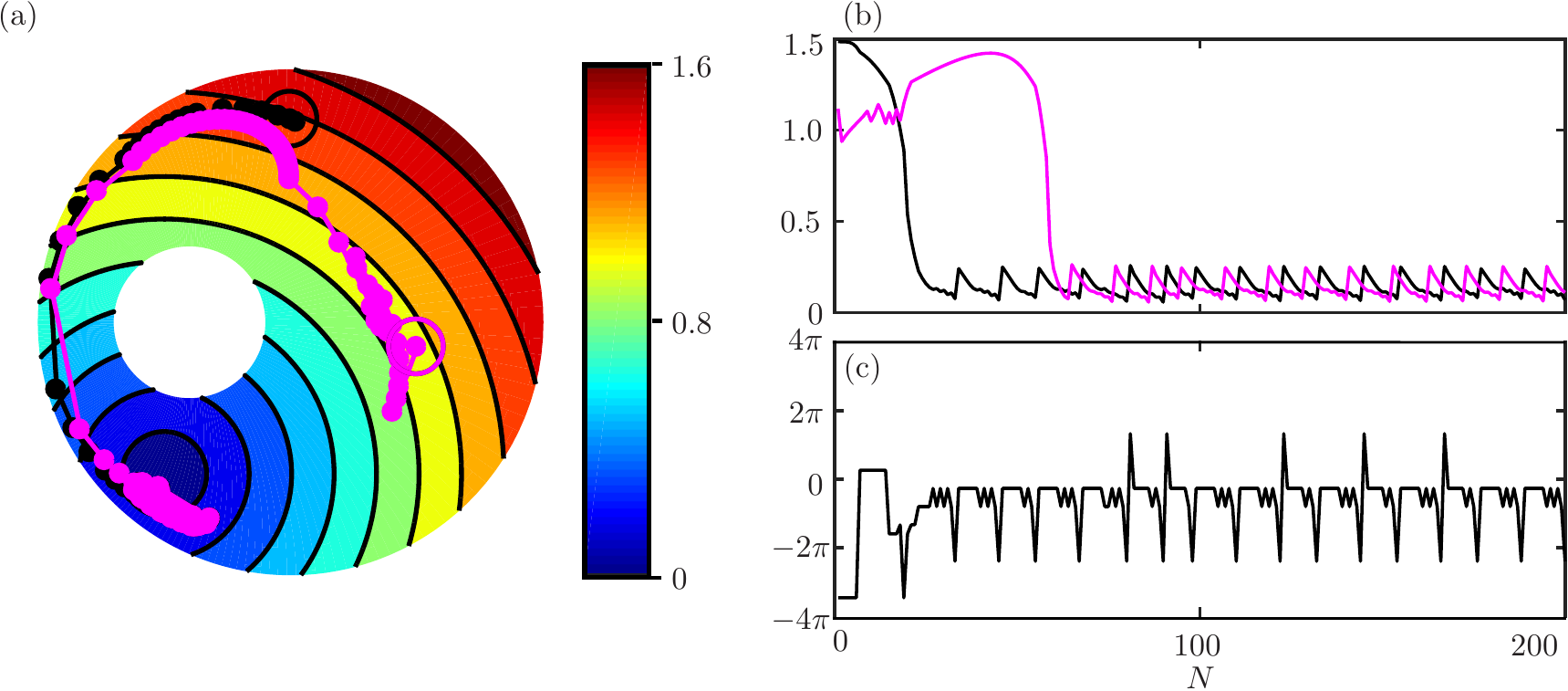}
\end{center}
\caption{Control of spheroidal particles.
(a) Contour levels of the distance from the target $(x_t,y_t)=(-0.5,-0.6)$. The trajectory of two particles departing from
 two different locations. The action chosen after each loop is done according with the 
Q-matrix obtained in the training stage. The position after each cycle corresponds with the black solid circles for a 
particle departing from $(x_0,y_0)=(0,0.8)$, whereas the particle departing from $(x_0,y_0)=(0.5,-0.1)$ is represented in magenta. 
(b) The evolution 
of the distance to the target as a function of the number of cycles for two trajectories represented in (a) with the same colour 
scheme. (c) The action chosen after each loop (the rotation of the angle of the cylinders) after each loop for the particle departing 
from $(x_0,y_0)=(0,0.8)$. }
\label{fig:fig6}
\end{figure*}

As the geometric phase is able to induce an effective displacement in the position and the angle of a particle after each loop, it constitutes a useful tool for controlling the spatial distribution of particles in bounded flows. Here we show how this effective displacement can be used to move a particle from an arbitrary initial position towards a given target by means of machine learning algorithms. In particular, we implemented a reinforcement Q-algorithm technique \cite{Watkins1992} that has been successfully used to control the motion of inertial particles in flows \cite{Colabrese2018}, to establish optimal protocols for geometric phase based displacements. The algorithm divides the journal-bearing flow into $N_s$ different states (depending on the distance to the target position) and drives the particle towards the target position $(x_T,y_T)$ by performing a sequence of $N_a$ closed loops actions.

The learning process consists of two stages: first, the particle is trained to explore the domain by performing sequences of actions that distinctly change the area of the parameter's loop in the $\in[-4\pi,4\pi]$ range. Each time that the particle changes its state for a given action, a reward inversely proportional to the square of the euclidean distance from the particle to the target is assigned to that action. If a particle advances closer to the target, the reward increases accordingly and the coefficients of the Q-matrix (of size $N_a\times N_s$) are updated. Consecutive actions are defined by choosing the action that, for the updated state, has the maximum Q-value (i.e., largest reward) to ensure it would bring the particle closer to the target. In a second stage, and after the training process is complete, the resulting Q-matrix is applied to move particles that depart from random initial position to the designated target. 

We performed $\sim 10^4$ training events departing from random initial positions within the domain and consisting of a total of 50 changes of state. In cases where particles converge and get stuck in a single state, we selected a random action after a small number of futile loops (e.g., 10)  to facilitate exploration of space but without updating the coefficients of the Q-matrix. Similarly, we always initialized all the elements of the  Q-matrix with a large value to ensure optimal probing of the actions space. We illustrate the resulting dynamics in Figure~\ref{fig:fig6}. Panel (a) shows two examples of trajectories approaching a selected target at $(x_T,y_T)=(-0.5,-0.6)$. To assess the robustness of the algorithm we did not stop the integration once the target had been reached but, instead, allow for a back and forth oscillation close to the target resulting from a continuous hopping between elements of the Q-matrix. This can be seen in the evolution of the distance to the target in Fig.~\ref{fig:fig6}(b) and in the sequence of selected actions in Fig.~\ref{fig:fig6}(c).

\section{Conclusions}

Although we have only illustrated the dynamics of spheroids under the particularly simple example of a two-dimensional journal-bearing flow, the basic principles of particle's rotations triggered by geometric phases we have described herein extend to more complex time dependent flows. Some possible applications of this work include the reorientation dynamics of fusiform planktonic microorganisms (e.g., pennate diatoms) in the upper layer of the ocean under the action of waves \cite{DiBenedetto2018}, bubbles in a fluid, that depending on the Bond / Eotvos number can be ellipsoidal in form\cite{Clift2013} (e.g., hair gels sold in transparent tubes with prominent bubbles are a familiar example, and bubbles in foods such as batters, souffl\'es and so on are important to their character \cite{campbell2016}) or the emergence of polar or nematic order in suspensions of spheroidal particles through the combination of a geometric phase and collective long-range hydrodynamics. Finally, reinforcement learning for the geometric phase could also be extended to other constrained loops (for instance, to peristaltic flows), wherever optimal strategies to drive particles to targets are relevant.

\section{Acknowledgements}
We acknowledge the support from the Spanish Ministry of Economy and Competitiveness (AEI, FEDER EU) grant nos FIS2016-77692-C2-1-P (I.T., O.P. and J.A.), IED2019-000958-I (IT), FIS2016-77692-C2-2-P (J.H.E.C.) and CTM- 2017-83774-D (J.A.). J.A thanks the Govern de les Illes Balears for financial support through the Vicenç Mut subprogram partially funded by the European Social Fund.

\bibliographystyle{aipnum4-1}

\end{document}